\documentstyle[pra,epsf,aps,twocolumn]{revtex}
\def\lfrac#1#2{#1/#2}
\def\sbf#1{{\footnotesize {\bf #1}}}

\begin{document}
\sloppy
\title{Second Topological Moment $\langle m^2 \rangle$
of Two Closed Entangled Polymers}
\author{Franco Ferrari$^{(1)}$
Hagen Kleinert$^{(2)}$
and Ignazio Lazzizzera$^{(1)}$\\
{$^{(1)}$\it Dipartimento di Fisica, Universit\`a di Trento, I-38050 Povo,
Italy\\
and INFN, Gruppo Collegato di Trento.}\\
{$^{(2)}$\it Institut f\"ur Theoretische Physik,\\
Freie Universit\"at Berlin, Arnimallee 14, D-14195 Berlin, Germany.}}
\date{February 2000}
\maketitle
\begin{abstract}
We calculate exactly by field theoretical techniques the second topological
moment
$\langle m^2 \rangle$ of entanglement
of two closed polymers $P_1$ and $P_2$.
This result is used to estimate approximately the mean square average of the
linking number
of a polymer $P_1$ in solution with other polymers.
\end{abstract}
\begin{figure}[tbhp]
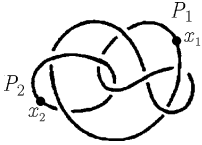

\input link.tps
\caption[]{Closed polymers $P_1, P_2$ with trajectories $C_1, C_2$
respectively.}
\label{Fig. 1}\end{figure}

{\bf 1.}
Consider two closed polymers  $P_1$ and $P_2$
which statistically can be linked with each other any  number of
times $m=0,1,2,\dots~$.
The situation is illustrated in Fig.~\ref{Fig. 1}
for $m= 2$.
An important physical quantity is the probability
distribution of the linking number $m$ as a function of
the lengths of $P_1$ and $P_2$.
As a first step towards finding it
we calculate,
in
this note,  an exact expression for the second moment of the distribution,
$\langle m^2 \rangle$.

An approximate result for this quantity was obtained before
in Ref.~\cite{BS} on the basis of a
a mean-field method, considering
the density of bond vectors of $P_2$
as Gaussian random variables. Such methods are usually quite accurate
when a large
number of polymers is involved \cite{BV,OV}. As an unpleasant feature, however, they
they introduce a dependence on the source of
Gaussian noise, and modify the critical behavior of the system, whereas
topological interactions are not expected to do that \cite{MK,ffil}.
Our note goes therefore an important step beyond this approximation.
It treats the two-polymer problem exactly, and contains
an application to the
topological entanglement in diluted solutions.
The relevance  of the two-polymer systems
to such systems
was emphasized
in \cite{BS}. Focusing attention upon a particular molecule,
$P_1$,
 one may imagine all others to form
 a single long effective molecule $P_2$.

{\bf 2.}
Let
$G_m ({\bf x}_1, {\bf x}_2; L_1, L_2)$ be
the configurational probability to
find the polymer $P_1$ of length $L_1$
with fixed coinciding end points at ${ \bf x}_1$
and
the polymer $P_2$ of length $L_2$
with fixed coinciding end points at ${ \bf x}_2$,
topologically entangled with
a Gaussian linking number $ m$.

The second moment
 $\langle m^2 \rangle$ is defined by the ratio of integrals \cite{tan}
\begin{eqnarray}
  \langle m^2 \rangle = \frac{\int d^3 x_1 d^3 x_2
     \int^{+\infty}_{-\infty } dm ~m^2 G_m
\left({\bf x}_1, {\bf x}_2; L_1, L_2\right)}
 {\int d^3 x_1d^3 x_2 \int^{+\infty}_{-\infty} dm\, G_m
    \left( {\bf x}_1, {\bf x}_2; L_1, L_2\right) }
\label{1}\end{eqnarray}
%

The denominator in (\ref{1}) plays the role of a partition function:
\begin{eqnarray}
Z\equiv
 {\int d^3 x_1 d^3 x_2 \int^{+\infty}_{-\infty} dm\, G_m
    \left( {\bf x}_1, {\bf x}_2; L_1, L_2\right) }.
\label{1Z}\end{eqnarray}

Due to the translational invariance of the system, the probabilities
depend only on the differences between the end point coordinates:
\begin{eqnarray}
 G_m \left( {\bf x}_1, {\bf x}_2; L_1, L_2\right) = G_m \left({\bf x}_1
 - {\bf x}_2;
L_1, L_2\right).
\label{2}\end{eqnarray}
Thus, after a shift of variables,
the spatial double integrals in (\ref{1}) can be
rewritten as
\begin{eqnarray}
\int\! d^3 x_1 d^3 x_2 G_m ({\bf x}_1
 \! - \!{\bf x}_2 ; L_1, L_2)
\!=\!
V\int\! d^3 x  G_m ({\bf x} ; L_1, L_2),   \nonumber
\label{3}\end{eqnarray}
where $V$ denotes the total volume of the system.

{\bf 3.}
The most efficient way of describing the statistical
fluctuations of the
polymers $P_1$ and $P_2$ is by
two complex polymer fields
$ \psi_1^{{ a}_1} ({\bf x}_1)$ and
 $\psi_2^{{ a}_2} ({\bf x}_2)$
with $n_1$ and $n_2$ replicas $(a_1=1,\dots,n_1,~a_2=1,\dots,n_2)$.
At the end we shall take $n_1,n_2\rightarrow 0$
to ensure that these  fields describe only one polymer each \cite{pi}.

For these fields we define an auxiliary
probability $G_ \lambda  (\vec {\bf x}_1, \vec{\bf x}_2 ;  \vec{z} )$
to find the polymer $P_1$
with open ends  at ${\bf x}_1,{\bf x}_1'$
and the polymer $P_2$
with open ends  at ${\bf x}_2,{\bf x}_2'$.
The double vectors
$\vec{{\bf x}}_1\equiv ({\bf x}_1,{\bf x}_1')$ and
$\vec{{\bf x}}_2\equiv ({\bf x}_2,{\bf x}'_2)$
collect initial and final
endpoints of the two polymers
  $P_1$ and $P_2$.
Here we follow the
approach of Edwards \cite{Ed}, in which one starts
with open polymers with fixed ends.
The case of closed polymers, where $m$ becomes a true topological number and
it is thus relevant in the present context,
is recovered in the limit of coinciding extrema.
We notice that in this way one introduces in the configurational probability
an artificial dependence on the fixed points
${\bf x}_1$ and
${\bf x}_2$. In physical
situations, however, the fluctuations of the polymers are entirely free.
For this reason we have averaged in (\ref{1}) over all possible
fixed points by means of the integrations in $d^3  {\bf x}_1d^3 {\bf x}_2 $.

The
auxiliary
probability $G_ \lambda  (\vec {\bf x}_1, \vec{\bf x}_2 ;  \vec{z} )$
is given
by a functional integral \cite{ffil}
\begin{eqnarray}
&&\!\!\!\!\!\!\!\!\!\!\!G_ \lambda  (\vec{{\bf x}}_1, \vec{{\bf x}}_2 ; \vec z
)
  =   \lim _{n_1, n_2 \rightarrow 0} \int
  {\cal D} ({\rm fields}) \,\nonumber \\
 & & \times \psi_1^{{a}_1} ({\bf x}_1)
  \psi_1^{* { a}_1}  ({\bf x}_1')
\psi_2^{{ a}_2} ({\bf x}_2) \psi_2^{*{ a}_2} ({\bf x}_2')
    e^{- {\cal A}},
\label{4}\end{eqnarray}
where
${\cal   D} (\mbox{fields})$
indicates the measure of functional
integration and $a_1,a_2$ are now fixed replica indices.
${\cal A}={\cal A}_{\rm CS}+{\cal A}_{\rm pol}$ is the action
governing the fluctuations.
It consists of a polymer action
\begin{eqnarray}
 {\cal A}_{\rm pol} =  \sum _{i=1}^{2} \int d^3{\bf x} \left[ |\bar {\bf D}^i
\Psi_i|^2 +
 m^2_i |\Psi_i |^2 \right].
\label{5}\end{eqnarray}
 and a  Chern-Simons action to
describe the linking number $m$
\begin{eqnarray}
 {\cal A}_{\rm CS}
 =i
{ \kappa }
\int d^3 x \,\varepsilon_{\mu \nu  \rho }
   A_1^\mu \partial _ \nu  A_2^\rho           ,
\label{5CS}\end{eqnarray}
In Eq. (\ref{5CS}) we have omitted a
gauge fixing term, which enforces the  Lorentz gauge.
The effects of self-entanglement and of
the so-called {\em excluded-volume\/} interactions
are ignored.
The Chern-Simons fields are coupled to the polymer fields by the
 covariant derivatives
\def\nablab{\BF \nabla}
\newcommand{\BF}[1]{\mbox{\boldmath $#1$}}
$
 {\bf D}^i = {\nablab} +i  \gamma _i {\bf A}^i,
$
with the coupling constants $ \gamma _{1,2}$
given by
$
  \gamma _1 =  \kappa,~
   \gamma _2 =  \lambda .$
The square masses of the polymer fields
are given by
$   m_i^2 = 2 M z_i$,
where
  $M=3/a$, with $a$ being the length of the polymer links,
and $z_i$
the chemical
potentials of the polymers,
 measured in units of the temperature.
The chemical
potentials  are  conjugate variables to the length
parameters $L_1$ and $L_2$, respectively.
The symbols $\Psi_i$ collect the replicas   $\psi_i^{a_i}$ of the
two polymer fields. Let us note that in the topological Landau-Ginzburg model
(\ref{4}) the
Chern-Simons fields do not change the critical behavior of the system,
as expected.

 The parameter $ \lambda $ is conjugate to the linking number $m$.
We can therefore calculate the desired probability
  $G_m (\vec {\bf x}_1, \vec {\bf x}_2; L_1, L_2)$
from the auxiliary one
  $G_ \lambda  (\vec {\bf x}_1, \vec {\bf x}_2; \vec z)$
by
 Laplace integrals over $\vec  z=(z_1,z_1)$ and an inverse
Fourier transformation
over $\lambda$.
%
%

{\bf 4.}
Let us use the polymer field theory
to calculate the partition function (\ref{1Z}).
 It is given by the integral
over the auxiliary probabilities
\begin{eqnarray}
 Z & = &\int d^3 x_1 d^3 x_2 \lim_{{\bf x}_1' \rightarrow {\bf x}_1
 \atop
   {\bf x}_2 '\rightarrow  {\bf x}_2}
   \int^{c + \infty}_{c - i \infty} \frac{Md z_1}{2\pi i}
   \frac{Mdz_2}{2\pi i} e^{z_1 L_1 + z_2 L_2} \nonumber \\
&&\times \int^{+ \infty}_{-\infty} d m \int^{+ \infty}_{-\infty} d \lambda
e^{-im \lambda }
     G_ \lambda \left( \vec{\bf x}_1 ,\vec{\bf x}_2 ; \vec z\right).
\label{15}\end{eqnarray}
  The integration over $m$ is trivial and gives
 $2  \pi   \delta ( \lambda )$, enforcing $ \lambda =0$, so that
\begin{eqnarray}
 Z & = & \int d^3 x_1 d^3 x_2 \lim_{{\bf x}_1 \rightarrow {\bf x}'_1
 \atop {\bf x}_2' \rightarrow {\bf x}_2}  \int^{c + i\infty}_{c - i \infty}
     \frac{Mdz_1}{2\pi i}\frac{Mdz_2}{2\pi i}
e^{z_1 L_1 + z_2 L_2} \nonumber \\
  & &~~~~~~~~~~~~~\times~~
    G_{ \lambda =0} \left( \vec{\bf x}_1 ,\vec{\bf x}_2 ; \vec z    \right).
\label{16}\end{eqnarray}
To calculate $
 G_{ \lambda =0} \left( \vec{\bf x}_1 ,\vec{\bf x}_2 ; \vec z
    \right)$ we observe
 that the action ${\cal A}$
  is quadratic
in $ \lambda $.
A trivial calculation gives
\begin{eqnarray}
 &&\!\!\!\!\!\!\!\!\!\!\!\!\!G_{ \lambda =0} \left( \vec{\bf x}_1 ,
\vec{\bf x}_2 ; \vec z
    \right)
   =  \int {\cal D} ({\rm fields}) e^{-{\cal A}_0} \nonumber \\
&& \times  \psi_1^{{a}_1}({\bf x}_1)
      \psi_1^{*{a}_1} ({\bf x}_1')\psi_2^{{a}_2} ({\bf x}_2) \psi_2^{{a}_2}
({\bf x}')
\label{22}\end{eqnarray}
where
\begin{eqnarray}
\!{\cal A}_0 &\equiv  &{\cal A}_{\rm CS}
\!+\! \!\int d^3 \!{\bf x}\! \left[ |{\bf D}_1 \Psi_1|^2\! +\! | \nablab \Psi_2
  |^2\! +\! \sum_{i =1} ^2  m_i^2|\Psi_i|^2 \right],
\label{18}
\end{eqnarray}
{}From Eq.~(\ref{18}) it is clear that
$G_{ \lambda =0} \left( \vec{\bf x}_1 ,\vec{\bf x}_2 ; \vec z
    \right)$
is the product of the configurational
probabilites of two free polymers.
In fact, the fields $\Psi_2, \Psi_2^*$ are free, whereas
the fields $\Psi_1, \Psi^*_1$ are apparently not free because
of the couplings with the Chern-Simons fields through the covariant
derivative ${\bf  D}^1$.
This is, however, an illusion:
Integrating out $A_2^\mu$ in (\ref{22}), we find
the flatness condition:
 $\varepsilon^{\mu \nu  \rho } \partial_ \nu  A^i_\mu= 0.$
 On a flat space with vanishing boundary conditions at infinity this
implies $ A_1^\mu  = 0$.
As a consequence,
the functional
integral (\ref{22}) factorizes
\begin{eqnarray}
\!\!\!
   G_{ \lambda =0} \left( \vec{\bf x}_1 ,\vec{\bf x}_2 ; \vec z    \right)
 = G_0 ({\bf x}_1 - {\bf x}_1'; z_1) \,G_0 ({\bf x}_2  - {\bf x}_2'
    ; z_2 )   ,
\label{24}\end{eqnarray}
where $ G_0 ({\bf x}_i - {\bf x}_i' ; z_i)$
are the free correlation functions of the polymer fields
\begin{eqnarray}
  G_0 ({\bf x}_i - {\bf x}_i'; L_i )
&=&\int^{c+i\infty}_{c-i\infty} \frac{Mdz_i}{2\pi i}e^{z_iL_i} G_0
   ({\bf x}_i - {\bf x}_i' ; z_i )
   \nonumber \\
 &  &\!\!\!\!\!\!\!\!\!\!\!\!\!\!\!\!\!\!\!\!\!\!\!\!\!\!\!\!\!\!\!\!\!
= \frac{1}{2 } \left(\frac{M}{4 \pi L_i}\right)^{3/2}
    e^{{- M({\bf x} _i -{\bf x}_i')^2}/{2L_i} }.
\label{27}\end{eqnarray}
Thus we obtain for  (\ref{16}) the integral
\begin{eqnarray}
 \!\!Z\! =\! 2  \pi \!\! \int\!\! d^3 x_1 d^3 x_2
 \lim_{{\bf x}_1' \rightarrow {\bf x}_1 \atop
                 {\bf x}_2'\rightarrow {\bf x}_2 }
  G_0 ({\bf x}_1 \!-\! {\bf x}_1' ; L_1 )\, G_0 ({\bf x}_2 \!-\! {\bf x}_2'
;L_2),
    \nonumber  \label{29}\end{eqnarray}
yielding the partition function
\begin{eqnarray}
Z =
   \frac{2 \pi  M^3 V^2}{(4 \pi )^3} (L_1 L_2)^{-3/2}.
   \label{30}\end{eqnarray}

It is important to realize  that in Eq.~(\ref{15})
 the limits of coinciding end points
${\bf x}_i' \rightarrow {\bf x}_i$ and
the inverse Laplace transformations
 do not commute unless a proper
renormalization scheme is chosen to eliminate the divergences
caused by the insertion of the composite operators $ |\psi_i^{a_i}
({\bf x})|^2$ and $|\Psi_i ({\bf x})|^2$.
%

\begin{figure}[tbhp]
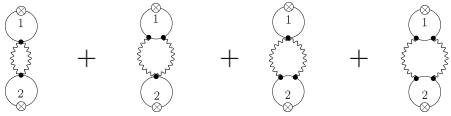

~\\[-2.5mm]{}~\input 2u4s.tps~\\
\caption[Four diagrams contributing in Eq.~(\protect\ref{39}).
 The lines indicate correlation functions of $\Psi_i$-fields.
The crossed circles with label $i$  denote the insertion of
$|\psi_i^{a_i}({\bf x}_i)|^2$]{Four diagrams
contributing in Eq.~(\protect\ref{39}).
 The lines indicate correlation functions of $\Psi_i$-fields.
The crossed circles with label $i$  denote the insertion of
$|\psi_i^{a_i}({\bf x}_i)|^2$.\label{4dia} }
\label{@FFF}\end{figure}

{\bf 5.}
Let us now turn to the numerator in Eq.~(\ref{1}). Exploiting
the identity $m^2e^{-im\lambda}=-{\partial^2}
e^{-im\lambda}/{\partial \lambda^2}$, and performing two partial integrations
 in $\lambda$,
the same technique
used above to evaluate the partition function $Z$ yields
\begin{eqnarray}
 N & = & \kappa ^2 \int d^3 x_1 d^3 x_2 \lim_{n_1 \rightarrow 0
\atop
    n_2 \rightarrow 0}
 \int^{c + i\infty}_{c - i \infty}  \frac{Md z_1 }{2\pi i}
      \frac{Mdz_2}{2\pi i} e^{z_1 L_1 + z_2 L_2}
   \nonumber \\
 &
    \times &\int {\cal D}(\mbox{fields})~    \exp (-{\cal A}_0)
\vert\psi_1^{a_1} ({\bf x}_1)\vert^2
     \vert \psi_2^{a_2}  ({\bf x}_2 ) \vert^2
 \nonumber \\
 &
\times& \left[  \left(\int d^3 x    \,{\bf A}_1 \cdot
 \Psi_1^*
{\bf \nablab} \Psi_1 \right)^2
 \!\!+ \frac{1}{2} \int d^3 x \, {\bf A}_1^2\,\vert \Psi_1 \vert^2
     \right]
   \nonumber \\
  &
\times& \left[  \left(\int d^3 x    \,{\bf A}_2 \cdot
 \Psi_2^*
{\bf \nablab} \Psi_2 \right)^2
\!\! + \frac{1}{2} \int d^3 x \, {\bf A}_2^2\,\vert \Psi_2 \vert^2
     \right].
\label{39}\end{eqnarray}
where ${\cal A}_0$ has been defined in (\ref{18}).
In the above equation we
have taken
the limits of coinciding endpoints
inside the
 Laplace integral over $z_1, z_2$.
This will be justified later
on the grounds that
the potentially dangerous Feynman diagrams containing
the insertions of operations like $|\Psi_i|^2$ vanish
in the limit $n_1, n_2 \rightarrow 0$.
The functional integral in
Eq.~(\ref{39})
can be calculated exactly
 by diagrammatic
methods since
only four diagrams shown in Fig.~(\ref{4dia})
contribute.%
Only the first diagram in Fig.~\ref{4dia}
is divergent from  the loop integral
formed by two correlation functions of the vector field.
This infinity may be absorbed in the four-$\Psi$
interaction
accounting for the
excluded volume effect which we do not consider at the moment.
No divergence arises from the
insertion of the composite fields
$\vert\psi_i^{a_i} ({\bf x}_i)\vert^2$.

{\bf 6.}
{}In this section we evaluate the first term appearing in the
right hand side of  Eq.~(\ref{39}):
\begin{eqnarray}
{}~N_1\!\! & = &  \frac{ \kappa ^2}{4}\lim_{n_1 \rightarrow 0 \atop n_2
\rightarrow 0}
 \int^{c + i\infty }_{c - i \infty} \frac{Md z_1}{2\pi i}
    \frac{Mdz_2}{2\pi i} e^{z_1 L_1 + z_2 L_2} \nonumber \\
 &&
           \int d^3
{x}_1  d^3 {x}_2
\int
     d^3 { x}_1' d^3 { x}_2'
  \label{40}\\
&& \bigg\langle \vert \psi_1^{a_1} ({\bf x}_1)\vert ^2
  \vert \psi_2^{a_2} ({\bf x}_2) \vert ^2
 \left(   \vert \Psi_1\vert^2 {\bf A}_1^2\right)_{{\sbf x}_1'}
\left(\vert \Psi_2\vert^2{{\bf A}_2^2}\right)_{{\sbf x}_2'}
 \bigg\rangle     . \nonumber
\end{eqnarray}
There is an ultraviolet-divergence which
must be
regularized.
This is done by cutting the spatial integrals off
at
the
persistence length $\xi$
over which a polymer is stiff.
This contains the stiffness caused by
 the excluded-volume effects. To be rigorous, we define
the integral (\ref{40})
on a lattice with spacing $\xi$.

Replacing the expectation values by the
Wick contractions corresponding to the first diagram in
Fig.~\ref{4dia},
 we obtain
\begin{eqnarray}
&&\!\!\!\!\!\!\!N_1 =\frac V{4\pi}\frac{M^4}{(4\pi)^6}(L_1L_2)^{-\frac
12}
\!\!\int_0^1\!\!ds\left[(1-s)s\right]^{-\frac 3 2}
 \!\!\int\!\! d^3xe^{-\frac {M{\sbf x}^2}{2s(1-s)}}\label{ioneint}\nonumber \\
&\times&\!
\int_0^1dt\left[(1-t)t\right]^{-\frac 3 2}
\!\int d^3ye^{- \frac{M{\sbf y}^2}{2t(1-t)}}\int
d^3x_1^{\prime\prime}\frac 1{\vert{\bf x}_1^{\prime\prime} \vert^4}  .
 \nonumber
\end{eqnarray}
%
The variables $\bf x$ and $\bf y$ have been rescaled with respect to
the original ones in order to extract the behavior of $N_1$ in $L_1$
and $L_2$. As a consequence, the lattices where $\bf x$ and $\bf
y$ are defined
have now spacings $ \xi/{\sqrt{L_1}}$ and $ \xi/{\sqrt{L_2}}$
respectively.
The ${\bf x},{\bf y}$ integrals may be explicitly computed by analytical
methods in the
physical limit $L_1,L_2>>\xi$, in which the above spacings become
small.
This has a physical explanation.
Indeed, if the polymer lengths are much larger than the persistence length,
the effects due to the finite monomer size become negligible and
can be ignored.

Finally, it is possible to approximate the integral in
${\bf x}_1^{\prime\prime} $ with an integral over a continuous
variable $\rho$ and a cutoff in the ultraviolet region:
$
\int d^3x_1^{\prime\prime}\lfrac{1}{{\vert{\bf x}_1^{\prime\prime}
\vert^4}}\sim
4\pi^2\int_\xi^\infty\lfrac{d\rho}{\rho^2}\label{appsone}.
 $
After these approximations, we obtain
\begin{eqnarray}
N_1 = {V   \sqrt{  \pi }} \frac{M}{(4 \pi )^3}
    (L_1 L_2)^{-1/2}  \xi^{-1}    .
\label{41}\end{eqnarray}

{\bf 7.}
For the second
diagram in Fig.~\ref{@FFF}
 we have to calculate
\begin{eqnarray}
&&
\!\!\!\!\!\!\!\!\!
\!\!\!\!\!\!\!\!\!\!\!\!\!
N_2
 =
 \kappa ^2  \lim_{n_1 \rightarrow 0 \atop n_2 \rightarrow 0}
     \int^{c + i \infty}_{c - i \infty} \frac{Mdz_1}{2\pi i}
\frac{Mdz_2}{2\pi i}
     e^{z_1 L_1 + z_2 L_2}
 \nonumber \\
 &&
         \!\!\!\!\!\!\!\!\!\!\!\!\!\times   \int d^3
{ x}_1  d^3 { x}_2
\int
     d^3 {x}_1'
d^3 {x}_1''
d^3 {x}_2'
\label{42}
\nonumber \\&&
 \!\!\!\!\!\!\!\!\!\!\!\!\!\times  \bigg\langle \vert \psi_1^{a_1}
({\bf x}_1)\vert^2
 \vert \psi_2^{a_2} ({\bf x}_2) \vert^2
 \left(
    \,{\bf A}_1 \cdot
 \Psi_1^*
{\bf \nablab} \Psi_1 \right)_{\sbf{x}_1'}
\nonumber \\
&& \!\!\!\!\!\!\!\!\!\!\!\!\!\times
 \left(
    \,{\bf A}_1 \cdot
 \Psi_1^*
{\bf \nablab} \Psi_1 \right)_{\sbf{x}_1''}
 \left( {\bf A}_2^2\,\vert \Psi_2 \vert^2\right)_{{\sbf x}_2'}
\bigg\rangle.
\label{@56}\end{eqnarray}
The above amplitude has no ultraviolet divergence, so that no regularization is
required.
The Wick contractions pictured in the second Feynman diagrams
of Fig.~\ref{4dia}
lead to the integral
\begin{eqnarray}
 N_2 = -4\sqrt{2} V L_2^{-1/2} L_1^{-1} \frac{M^3}{\pi^6} \int_{0}^{1}
 dt \int^{t}_{0} dt' C(t,t'),
\label{43}
\label{@57}
\end{eqnarray}
 where $C(t,t')$ is a function independent of $L_1$ and $L_2$:
\begin{eqnarray}
  &&C(t,t')  =  \left[(1-t) t' (t-t') \right] ^{-3/2}
\int
     d^3 { x} d^3 { y}  d^3 { z} e^{-\frac{ M ({\sbf y} - {\sbf x })^2 }{
     2(1-t)}}
 \nonumber \\ && \times
 \left( \nabla^ \nu_{\sbf y} e^{-M {\sbf y}^2 /2t'}\right) \left(
     \nabla^\mu_{\sbf x}  e^{-M {\sbf x}^2 /2 (t-t') }\right)
P_{\mu \nu }({\bf x},{\bf y},{\bf x}),
\label{44}
\label{@58}         \nonumber
\end{eqnarray}
with
$P_{\mu \nu }({\bf x},,{\bf y},{\bf x})\equiv
[ \delta _{\mu  \nu } {\bf z} \cdot
     ({\bf z} + {\bf x}) - \left({ z} + { x}\right)_\mu
  { z}_ \nu ]/
   ({\vert{\bf z}\vert^3 \vert {\bf z} + {\bf x}\vert^3})$.
As in the previous section, the variables $\bf x, \bf y, \bf z$ have been
rescaled with respect to the original ones in order
to extract the behavior in $L_1$.
Again, if $L_1,L_2>>\xi$
the analytical evaluation of
$C(t,t')$ becomes possible, leading to 
\begin{eqnarray}
  N_2 & = & - \frac{V L_2^{-1/2}L_1^{-1} }{(2\pi)^6}
 M^{3/2}   4 K,
\label{46}\end{eqnarray}
where $K$ is the constant
$ \frac{1}{6} B \left(\frac{3}{2},\frac{1}{2}\right) +
 \frac{1}{2} B\left(\frac{5}{2}, \frac{1}{2}\right)
- B
 \left(\frac{7}{2}, \frac{1}{2}\right) + \frac{1}{3}
 B \left(\frac{9}{2}, \frac{1}{2}\right)   ={19\pi}/{384} \approx0.154
		 ,$
and $B(a,b)$ is the Beta function.
For large  $L_1 \rightarrow \infty $, this diagram gives a negligible
contribution with respect to $N_1$.

The third diagram in Fig.~\ref{4dia}
give the same as the second, except that $L_1$ and $L_2$ are interchanged:
$
N_3=N_2|_{L_1\leftrightarrow  L_2}.
$

{\bf 8.}
The fourth Feynman diagram in
Fig.~\ref{@FFF}
has no ultraviolet divergence.
As before, it can be exactly evaluated apart from the lattice integrations.
However, the behavior of the related
Feynman integral $N_4$
can be
easily estimated in the following limits:

1. $L_1 \gg 1; L_1 \gg L_2$, where
$N_4 \propto L_1^{-1},$

2. $L_2 \gg 1; L_2 \gg L_1$ where
 $N_4  \propto L_2^{-1},$

3. $L_1, L_2 \gg 1,~ {L_2}/{L_1} =  \mbox{finite} ,$ where
  $N_4 \propto L_1^{-3/2}$.\\
Moreover, if the lengths of the polymers are considerably larger than
the
persistence length, $N_4$ can be computed in a
closed form:
%
%
\begin{eqnarray}
 N_4 &  \approx & - \frac{128V}{\pi^5} \frac{M}{\pi^{3/2}}
 (L_1L_2)^{-1/2}
 \nonumber \\
& \times &\int^{1}_{0} ds \int^{1}_{0} dt
    (1-s) (1-t) (st)^{1/2} \nonumber \\
&\times &\left[ L_1 t (1-s) + L_2 (1-t)s\right] ^{-1/2} .
  \label{57}\end{eqnarray}
It  is simple to check that this expression has exactly
the above
behaviors.

{\bf 9.}
Collecting all contributions we obtain the result for the
second topological moment
$ \langle m^2\rangle = \lfrac{(N_1 + N_2 + N_3 +  N_4)}{Z} ,$
with $N_1,\,N_2,\,N_3,\,N_4,\,Z$
given by Eqs.~(\ref{30}),
(\ref{41}),
(\ref{46}),
and
(\ref{57}).
In all formulas, we have assumed that
 the volume $V$ of the system is much larger
than the size of the volume occupied by a single polymer,
i.e.,
$V \gg L_1^3$.

To discuss the physical content of the above expression for
$\langle m^2\rangle$, we consider a
number $N_p$ of polymers $p_1\ldots p_{N_p}$ with lengths
$l_1\ldots l_{N_p}$ in an uniform solution.
We introduce the polymer concentration
$\rho= \lfrac{{\cal M}}{V}
 $ as the average mass density  of the polymers per unit volume,
where ${\cal M}$ is the total mass of the polymers
 ${\cal M} = \sum_{k = 1}^{N_p} m_a \lfrac{l_k}{a}$ and
$m_a $ is  the mass of a single monomer of length $a$.
Thus ${l_k}/{a}$ is the  number of monomers in the
polymer $p_k$.
The polymer $P_1$ is singled out as anyone of the polymers $p_k$, say $p_{\bar
k}$,
of length $L_1=l_{\bar k}$.
The
remaining ones are replaced by a long effective polymer  $P_2$
of length $L_2= \sum_{k\neq \bar k} l_k$.
{}From the above relations we may also write
\begin{eqnarray}
  L_2 \approx \frac{a V \rho }{m_a}.
\label{69}\end{eqnarray}
In this way, the length of the effective molecule
$P_2$ is expressed in terms of physical parameters.
Keeping only the
leading terms for $V \gg 1$,
we find for
 the average square number of intersections $\langle m^2\rangle_{sol}$
formed by $P_1$ with
the other polymers
the approximate result
\begin{eqnarray}
 \langle m^2 \rangle_{sol} \approx \frac{N_1 + N_2}{Z},
\label{70}\end{eqnarray}
which, in turn, has the approximate form
\begin{eqnarray}
\langle m^2 \rangle_{sol} = \frac{a\rho }{m_a} \left[
   \frac{\xi^{-1} L_1}{2 \pi ^{1/2} M^2}-
     \frac{2K L_1^{1/2} }{\pi^4 M^{3/2}} \right],
\label{71}\end{eqnarray}
with $K$ as defined after Eq.~(\ref{46}).
This is the announced final result. Since the persistence length is of the same
order of the monomer length $a$ and $M\sim a^{-1}$,
 $\langle m^2 \rangle$ is positive for large $L_1$ as it should.

{\bf 10.}
In conclusion, we have found an exact
field theoretic formula
for the second topological moment of two polymers.
Only the
final
 integrations over the spatial variables in the
Feynman diagrams of Fig.~\ref{4dia} were done approximately.
These were defined on a lattice related to the finite monomer size.
Our  Chern-Simons-based theory is free of
the shortcomings
of previous  mean-field procedures.
Our formula for $\langle m^2\rangle$
has been applied to the realistic case of long flexible polymers
in a solution. When the polymer lengths become large,
the Feynman integrals in $N_1,\ldots,N_4$
can be
evaluated
analytically.
In this way we have been able to derive
the result (\ref{71}) for the average square number
of intersections formed by a polymer $P_1$ with all the others.
This calculation is
{\em exact\/} in the long-polymer limit.
The corrections
to (\ref{71}) are
suppressed by
further inverse  square roots of the polymer lengths.

To leading order in $L_1$, our result (\ref{71})
agrees
 with that of \cite{BSII}, but our exact subleading
correction go beyond
the approximation of
\cite{BSII}.
Note that there is no direct comparison of our result
with that of \cite{tan},
since there the  polymer $P_2$ was considered as a fixed  obstacle
causing a dependence on the choice of the
configuration of  $P_2$.

Finally, let us emphasize
the absence of infrared divergences
in the topological field theory
(\ref{4}) in the limit of vanishing masses $m_1,m_2=0$.
As a consequence,
the second topological moment does not diverge in the limit of
large $L_1$
if
$\langle m^2 \rangle$
is calculated from (\ref{4})
for polymers passing through
two fixed points
${\bf x}_1,{\bf x}_2 $.
This indicates a much stronger
reduction of the
configurational fluctuations by topological
constraints than
one might have anticipated.

\end{document}